\documentclass{article}

\usepackage{authblk}
\usepackage[font=small]{caption}
\usepackage[utf8]{inputenc}
\usepackage[english]{babel}

\usepackage{amsmath,bm, amsfonts,amssymb,amsthm}
\usepackage{multirow}
\usepackage{footnote}
\usepackage{hyperref}
\usepackage[square,sort,comma,numbers]{natbib}
\usepackage{hypernat}
\usepackage{graphicx}

\DeclareMathOperator*{\argmin}{arg\,min}

\let\citeleft=(
\let\citeright=)

\begin{document}
	
	\pdfinfo{
		/Author (AUTHORS)
		/Title (TITLE)
	}
	
	\title{\vspace{-2cm}Model-Based Reconstruction for Simultaneous Multi-Slice T1 Mapping using Single-Shot Inversion-Recovery Radial FLASH}
	
	\author[1,2]{Xiaoqing Wang}
	\author[1,2]{Sebastian Rosenzweig}
	\author[1,2]{Nick Scholand}
	\author[1,2]{H. Christian M. Holme}
	\author[1,2,3,4]{Martin Uecker}
	
	\affil[1]{\small Institute for Diagnostic and Interventional Radiology of the University Medical Center G\"ottingen, Germany}
	\affil[2]{\small German Centre for Cardiovascular Research (DZHK), Partner Site G\"ottingen, Germany}
	\affil[3]{\small Cluster of Excellence ``Multiscale Bioimaging: from Molecular Machines to Networks of Excitable Cells'' (MBExC), University of G\"ottingen, Germany}
	\affil[4]{\small Campus Institute Data Science (CIDAS), University of G\"ottingen, Germany}
	\maketitle
	
	\vfill
	\noindent
	\textit{Running head:} Model-based reconstruction for SMS T1 mapping
	
	\noindent
	\textit{Address correspondence to:} \\
	Dr.rer.nat. Xiaoqing Wang, Institute for Diagnostic and Interventonal Radiology of the University Medical Center G\"ottingen,
	Robert-Koch-Str. 40, 37075 G\"ottingen, Germany \\
	xiaoqing.wang@med.uni-goettingen.de

	\noindent
	This work was supported by the DZHK (German Centre for Cardiovascular Research),
	by the Deutsche Forschungsgemeinschaft (DFG, German Research Foundation) under
	Germany’s Excellence Strategy - EXC 2067/1- 390729940, and funded in part
	by NIH under grant U24EB029240.
	
	\noindent
	Approximate word count: 237 (Abstract) 4171 (body)\\
	
	\noindent
	Submitted to \textit{Magnetic Resonance in Medicine} as a Full Paper.
	
	\clearpage
	
	\section*{Abstract}
	
	\noindent \textbf{Purpose}: To develop a single-shot multi-slice T1 mapping method by combing simultaneous multi-slice (SMS) excitations, single-shot inversion-recovery (IR) radial fast low-angle shot (FLASH) and a nonlinear model-based reconstruction method.
	
	\noindent \textbf{Methods}: SMS excitations are combined with a single-shot IR radial FLASH sequence for data acquisition. A previously developed single-slice calibrationless model-based reconstruction is extended to SMS, formulating the estimation of parameter maps and coil sensitivities from all slices as a single nonlinear inverse problem. Joint-sparsity constraints are further applied to the parameter maps to improve T1 precision. Validations of the proposed method are performed for a phantom and for the human brain and liver in six healthy adult subjects.

	\noindent
	\textbf{Results}:  Phantom results confirm good T1 accuracy and precision of the simultaneously acquired multi-slice T1 maps in comparison to single-slice references. In-vivo human brain studies demonstrate the better performance of SMS acquisitions compared to the conventional spoke-interleaved multi-slice acquisition using model-based reconstruction. Apart from good accuracy and precision, the results of six healthy subjects in both brain and abdominal studies confirm good repeatability between scan and re-scans. The proposed method can simultaneously acquire T1 maps for five slices of a human brain ($0.75 \times 0.75 \times 5$ mm$^3$) or three slices of the abdomen ($1.25 \times 1.25 \times 6$ mm$^3$) within four seconds.

	\noindent \textbf{Conclusion}: The IR SMS radial FLASH acquisition together with a non-linear model-based reconstruction enable rapid high-resolution multi-slice T1 mapping with good accuracy, precision, and repeatability.
	
	\noindent
	\textbf{Keywords}: simultaneous multi-slice, model-based reconstruction, T1 mapping, radial FLASH
	
	\clearpage

	\section*{Introduction}
	\label{sec:introduction}

	Quantitative mapping of MR relaxation times such as T1 finds increasing
	applications in a variety of clinical use cases \cite{margaret2012practical,
		kellman2014t1}. Mapping of T1 relaxation time commonly relies on the inversion-recovery (IR)
	Look-Locker sequence where RF excitations are continuously
	applied after inversion followed by computation of a T1 map in a postprocessing step
	\cite{Look_Rev.Sci.Instrum._1970,deichmann1992quantification,Messroghli_Magn.Reson.Med._2004}.
	While efficient, the conventional IR Look-Locker method may still require segmented data
	acquisitions with multiple inversions
	\cite{deichmann1992quantification,Messroghli_Magn.Reson.Med._2004}.  As a
	sufficient delay is necessary between inversions, these techniques still suffer
	from long measurement time. Most recently, advances in sequence development
	such as non-Cartesian sampling together with state-of-the-art reconstruction
	techniques have enabled faster parameter mapping
	\cite{Block_IEEETrans.Med.Imaging_2009, Doneva_Magn.Reson.Med._2010,
		Petzschner_Magn.Reson.Med_2011,huang2012t2,velikina2013accelerating,
		zhang2015accelerating,Tamir_Magn.Reson.Med._2017,feng2017compressed},
	including accelerated T1 mapping within a
	single inversion recovery \cite{gensler2014myocardial,
		wang2016high,marty2018fast}. These methods usually consist of two steps: First,
	reconstruction of contrast-weighted images from undersampled datasets and,
	second, subsequent voxel-by-voxel fitting of the T1 map. In contrast,
	nonlinear model-based reconstruction methods
	\cite{Block_IEEETrans.Med.Imaging_2009,Fessler_IEEESignalProcess.Mag._2010,
		Sumpf_J.Magn.Reson.Imaging_2011,tran2013model,zhao2014model,peng2014exploiting,knoll2015model,
		Volkert_Int.J.Imag.Syst.Tech._2016,zhao2016maximum, Wang_Magn.Reson.Med._2018,
		Maier2019} estimate parameter maps directly from k-space, completely bypassing
	the intermediate step of image reconstruction and voxel-by-voxel fitting. Moreover,
	a~priori information such as sparsity constraints can be applied to the
	parameter maps to improve precision
	\cite{Block_IEEETrans.Med.Imaging_2009,zhao2014model,knoll2015model,Wang_Magn.Reson.Med._2018}.

	So far, most of the above efforts have focused on the acceleration of
	single-slice parameter mapping.  However, in clinical applications, multi-slice
	parameter mapping is highly desirable. For example, it has been recommended to
	perform myocardial T1 mapping in at least three short-axis sections to capture
	potential heterogeneity across the left ventricular wall
	\cite{moon2013myocardial,weingartner2017simultaneous}.  Methods exploiting the
	conventional multi-slice acquisition strategy have been reported
	\cite{shah2001new, deichmann2005fast, Wang2018,li2019rapid}. On the other
	hand, the simultaneous multi-slice (SMS) technique
	\cite{Barth_Magn.Reson.Med._2016} is a promising way to accelerate multi-slice
	quantitative MRI. SMS allows for the distribution of undersampling along the
	additional slice dimension and exploits sensitivity encoding in all three
	spatial dimensions. Applications of SMS in quantitative MRI include but are not
	limited to simultaneous three-slice MR fingerprinting \cite{ye2017simultaneous},
	simultaneous three-slice cardiac T1 mapping \cite{weingartner2017simultaneous}
	based on the SAPPHIRE technique and simultaneous multi-slice T2 mapping using
	the Cartesian multi-echo spin-echo sequence with a model-based iterative
	reconstruction \cite{hilbert2019fast}.

	To further enable fast T1 mapping of multiple slices, in this work, we aim to
	combine simultaneous multi-slice excitations and single-shot IR radial FLASH
	with a nonlinear model-based reconstruction to enable multi-slice T1 mapping
	within a single inversion recovery. In particular, we first combine SMS
	excitation with the single-shot IR radial FLASH sequence using a golden-angle
	readout. Next, we extend a previously developed single-slice calibrationless
	model-based reconstruction to SMS, formulating the estimation of parameter maps
	and coil sensitivities from all slices as a single nonlinear inverse problem.
	In this way, no additional coil-calibration steps are needed. Further,
	joint-sparsity constraints are applied on the parameter maps to improve T1
	precision. Performance of the proposed method is validated first on an
	experimental phantom and then on human brain and liver studies of six
	healthy adult subjects.

	\section*{Methods}
	\label{sec:theory}
	\subsection*{Sequence Design}
	
	The sequence starts with a non-selective adiabatic inversion pulse, followed by
	a radial FLASH readout using continuous SMS excitations and a tiny golden angle
	between successive spokes ($\approx 23.63^{\circ}$) \cite{wundrak2016golden}.
	In the partition dimension, radial spokes can be designed to follow an aligned
	or non-aligned distribution as shown in Figure 1. The aligned scenario allows
	decoupling of SMS data using an
	inverse Fourier transform along the partition dimension, followed by
	independent reconstruction of each slice. In such a way, there is still the
	SNR benefit of a SMS acquisition over the spoke-interleaved multi-slice scheme.
	However, the main advantage of SMS - acceleration in the direction perpendicular
	to the slices - only comes into play when distinct k-space samples are acquired
	in each partition \cite{zhou2017golden, Rosenzweig_Magn.Reson.Med._2018}.
	Figure 1 (bottom) depicts the latter case, where a larger k-space coverage could
	be achieved within a given readout time. In this work, we applied the tiny golden
	angle ($\approx 23.63^{\circ}$) sampling in the partition dimension as well.
	Accordingly, throughout this manuscript, we adopt the terms "SMS aligned" and
	"SMS golden-angle" to refer to the two SMS acquisition strategies, respectively.
	To reconstruct images/parameter maps in the latter SMS golden-angle acquisition
	scheme, conventional slice-by-slice reconstruction can not be used. A more
	general SMS model-based reconstruction method is therefore developed, which is
	explained in the following.

	\subsection*{SMS Model-based Reconstruction}
	
	Following a similar notation introduced in
	\cite{Rosenzweig_Magn.Reson.Med._2018}, we define $p, q \in \{1,\ldots,Q\}$ as the
	partition index and the slice index, respectively, where $Q$ is the total number of
	partitions/slices. In SMS acquisitions, the signal from the $p$th partition
	$\widetilde{\bm{y}}^{p}$ can be written as
	\begin{equation}
	\label{signal}
	\widetilde{\bm{y}}^{p} = \sum_{q=1}^{Q}{\xi}^{p,q} \bm{y}^{q},
	\end{equation}
	where ${\xi}^{p,q}$ is an SMS encoding matrix. In this study, it is chosen as the
	Fourier matrix, i.e., ${\xi}^{p,q} = \exp\Big(-2\pi i\frac{(p-1)(q-1)}{Q}\Big)$.
	The signal $\bm{y}_{j}^{q}(t)$ for the $q$th slice of the $j$th coil is given by
	\begin{equation}
	\bm{y}_{j}^{q}(t) = \int M^{q}(\vec{r})c^{q}_{j}(\vec{r})e^{-i\vec{r}\cdot\vec{k}(t)}d\vec{r},
	\label{eq2}
	\end{equation}
	where $c^{q}_{j}$ is the corresponding coil sensitivity map, $\vec{r}$ is the
	position in image space, and $\vec{k}(t)$ is the chosen k-space trajectory. $M^{q}$
	is the $T_{1}$ relaxation model for the $q$th slice at the inversion time
	$t_{k}$:
	\begin{equation}
	\label{T1_relax}
	M_{t_{k}}^{q} = M_{ss}^{q} - (M_{ss}^{q} + M_{0}^{q}) \cdot e^{-t_{k}\cdot {R^{*}_{1}}^q},
	\end{equation}
	where $t_{k}$ is defined as the center of each acquisition window.
	$M^{q}_{ss}$, $M^{q}_{0}$ and ${R^{*}_{1}}^q$ are the steady-state signal,
	equilibrium signal and effective relaxation rate, respectively. After
	estimation of ($M_{ss}^{q}$, $M_{0}^{q}$, ${R^{*}_{1}}^{q}$), $T_{1}$ values of
	the $q$th slice can be calculated by
	\cite{Look_Rev.Sci.Instrum._1970,deichmann1992quantification, deichmann2005fast}: $T_{1}^{q} =
	\frac{M_{0}^{q}}{M_{ss}^{q}\cdot {R^{*}_{1}}^{q}} + 2 \cdot \delta t$, with $\delta t$ the delay 
	between inversion and the start of data acquisition, which is around 15 ms for the sequence used in this study.
	To estimate parameter maps and coil sensitivity maps from all slices, equations
	(\ref{signal}) and (\ref{eq2}) are understood as a nonlinear inverse problem
	with a nonlinear operator $F$, mapping all the unknowns from all slices to the
	measured undersampled SMS data
	$\widetilde{Y} = (\widetilde{\bm{y}}_{t_{1}}^{1},
	\ldots,\widetilde{\bm{y}}_{t_{1}}^{Q},\widetilde{\bm{y}}_{t_{2}}^{1}, \ldots,
	\widetilde{\bm{y}}_{t_{n}}^{Q})^{T}$:
	
	\begin{equation}
	F:x \mapsto \bm{P}{\Xi}\left(
	\begin{array}{c}
	\mathcal{F}\{ \bm {c}^{1} \cdot \bm{M}_{t_{1}}^{1}\} \\
	\vdots \\
	\mathcal{F}\{ \bm {c}^{Q} \cdot \bm{M}_{t_{1}}^{Q}\} \\
	\mathcal{F}\{ \bm {c}^{1} \cdot \bm{M}_{t_{2}}^{1}\} \\
	\vdots \\
	\mathcal{F}\{ \bm{c}^{Q} \cdot \bm{M}_{t_{n}}^{Q}\}
	
	\end{array}\right),  \quad \text{with} \quad
	\bm{P}{\Xi} := 
	\left(
	\begin{array}{cccc}
	P_{t_{1}}{\xi} & && 0 \\
	&P_{t_{2}}{\xi} & & \\
	&  & \ddots &  \\
	0 & && P_{t_{n}}{\xi}  
	\end{array}
	\right)
	\end{equation}
	
	\begin{equation}
	\text{and} \quad
	\mathcal{F}\{\bm{c}^{q} \cdot \bm{M}_{t_{k}}^{q}\}:=\left(
	\begin{array}{c}
	\mathcal{F}\{  c^{q}_{1} \cdot M_{t_{k}}^{q}(M_{ss}, M_{0}, R_{1}^{*}) \}  \\
	\vdots \\
	\mathcal{F}\{c^{q}_{N} \cdot M_{t_{k}}^{q}(M_{ss}, M_{0}, R_{1}^{*}) \}  \\
	\end{array}\right).
	\end{equation}
	
	Here $\bm{P}$ is the sampling operator for the given k-space trajectory and
	$\mathcal{F}$ is the two dimensional Fourier transform. $\bm c^{q}$ represents
	a set of coil sensitivity maps for the $q$th slice. $M_{t_{k}}^{q}(\cdot)$ is
	the relaxation model described in Equation (\ref{T1_relax}).
	The unknowns are $x = (x^{1}, \cdots, x^{q},\cdots, x^{Q})^{T}$ with $x^{q} =
	(M^{q}_{ss}, M^{q}_{0}, {R^{*}_{1}}^q, c^{q}_{1},\cdots, c^{q}_{N})^{T}$, which
	are then estimated by solving the following regularized nonlinear inverse
	problem:
	\begin{equation}
	\hat{x} = \argmin_{x\in D} \|F(x) -\widetilde{Y} \|_{2}^{2} + \alpha \sum_{q=1}^{Q}R(x^{q}_{\bm{m}}) + \beta \sum_{q=1}^{Q}U(x^{q}_{\bm{c}}),
	\end{equation}
	with $R(\cdot)$ the joint $\ell_{1}$-Wavelet regularization in the parameter
	dimension \cite{Wang_Magn.Reson.Med._2018} and $U({\cdot})$ the Sobolev norm
	\cite{Uecker_Magn.Reson.Med._2008} enforcing the smoothness of coil sensitivity
	maps. $x^{q}_{\bm{m}} = (M^{q}_{ss}, M^{q}_{0}, {R^{*}_{1}}^q)^{T}$ and
	$x^{q}_{\bm{c}} = (c^{q}_{1},\cdots, c^{q}_{N})^{T}$, $\alpha$ and $\beta$ are
	the regularization parameters for the parameter and coil-sensitivity maps,
	respectively. $D$ is a convex set ensuring ${R^{*}_{1}}^q$ to be nonnegative.
	The above nonlinear inverse problem is then solved by the iteratively
	regularized Gauss-Newton method (IRGNM) where in each Gauss-Newton step the
	nonlinear problem is linearized and solved by the fast iterative
	shrinkage-thresholding algorithm (FISTA)
	\cite{Beck_IEEETrans.ImageProcessing_2009}. More details on the IRGNM-FISTA
	algorithm can be found in \cite{Wang_Magn.Reson.Med._2018}.
	
	\subsection*{Data Acquisition}
	
	All MRI experiments were conducted on a Magnetom Skyra 3T (Siemens
	Healthineers, Erlangen, Germany) with approval of the local ethics committee.
	The proposed method was first validated on a commercial reference phantom
	(Diagnostic Sonar LTD, Scotland, UK) consisting of six compartments with
	defined T1 values surrounded by water. Phantom and brain studies were conducted
	with a 20-channel head/neck coil, whereas abdominal scans were performed with a
	combined thorax and spine coil with 26 channels. Six subjects (4 females, 2
	males, 26 $\pm$ 5 years old) without known illness were recruited. For each
	subject, both brain and liver measurements were performed. In all experiments,
	simultaneously acquired slices using IR radial FLASH are separated by a fixed
	distance $d$. All single-slice and multi-slice acquisitions employed the same
	nominal flip angle $\alpha$ = $6^{\circ}$. Acquisition parameters for phantom
	and brain measurements were: FOV: $192 \times 192$ mm$^{2}$, matrix size: $256
	\times 256$, TR/TE = 4.10/2.58 ms, bandwidth 630 Hz/pixel, slice thickness
	$\Delta z$ = 5 mm, slice distances $d$ = 15 mm and $d$ = 20 mm for phantom and
	in-vivo studies, respectively. A gold standard T1 mapping was performed on the
	center slice of the phantom using an IR spin-echo method
	\cite{Barral_Magn.Reson.Med._2010} with 9 IR scans (TI = 30, 530, 1030, 1530,
	2030, 2530, 3030, 3530, 4030 ms), TR/TE = 4050/12 ms, FOV: 192 $\times$ 192
	mm$^{2}$, matrix size: $192\times 192$, and a total acquisition time of 2.4
	hours. Parameters for the abdominal measurements were: FOV: $320 \times 320$
	mm$^{2}$, matrix size: $256 \times 256$, TR/TE = 2.70/1.69 ms, bandwidth 850
	Hz/pixel, $\Delta z$  = 6 mm, $d$ = 20 mm. All single-shot measurements were
	acquired within a 4-second duration, which was chosen to compromise between good T1 accuracy, 
	SNR while still keep the acquisition time short \cite{wang2016high}. 
	The SMS golden-angle acquisitions were executed twice
	to evaluate the repeatability of the proposed method. Abdominal experiments
	were performed during a brief breathhold.
	
	\subsection*{Numerical Implementation}
	
	All SMS image reconstruction was done offline based on the software package
	BART \cite{Uecker__2015} using a 40-core 2.3 GHz Intel Xeon E5–2650 server with
	a RAM size of 512 GB. For abdominal studies, coil 
	elements far away from the regions of interest (i.e., coils near the arm regions)
	were manually excluded to image reconstruction to remove residual streaking
	artefacts \cite{Block_J.KoreanSoc.Magn.Reson.Med._2014}. After gradient-delay
	correction \cite{Block__2011} and
	channel compression to 8 principle components, the multi-coil radial raw data
	were gridded onto a Cartesian grid, where all successive iterations were
	then performed using FFT-based convolutions with the point-spread function
	\cite{Wajer__2001,Uecker_Magn.Reson.Med._2010}.  Parameter maps $(M^{q}_{ss}, M^{q}_{0},
	{R^{*}_{1}}^q)^{T}$ were initialized with $(1.0, 1.0, 2.0)^{T}$ and all coil
	sensitivities were initialized with zeros for all slices. The regularization parameters
	$\alpha$ and $\beta$ were initialized with $1.0$ and subsequently reduced by a
	factor of three in each Gauss-Newton step. A minimum value of $\alpha$ was used to
	control the noise of the estimated parameter maps even with a larger number
	of Gauss-Newton steps. The optimal value $\alpha_{\rm min}$ was chosen manually
	to optimize SNR without compromising the quantitative accuracy or delineation of
	structural details. 10 Gauss-Newton steps were employed to ensure convergence. 
	With these settings, it took around 10 hours to reconstruct a five-slice SMS brain
	dataset and 6 hours to reconstruct a three-slice SMS abdominal dataset on
	the CPU system. For single-slice references, reconstructions were able to run on
	a GPU (Tesla V100 SXM2, NVIDIA, Santa Clara, CA), which then took only
	8 to 12 min per dataset.
	
	Model-based reconstruction techniques generally offer a flexible choice of
	temporal binning, i.e., even a single radial spoke per k-space frame could
	be employed for accurate parameter estimation
	\cite{Volkert_Int.J.Imag.Syst.Tech._2016}. However, a certain amount of
	temporal binning effectively reduces the computational demand as long as the
	T1 accuracy is not compromised \cite{deichmann1992quantification,
		Volkert_Int.J.Imag.Syst.Tech._2016}. Here, the number of binned spokes was
	chosen such that the temporal bin size does not exceed 85 ms, which
	is sufficiently low to keep the quantification error below 1\% as
	shown previously \cite{Volkert_Int.J.Imag.Syst.Tech._2016}.
	More
	specifically, 15, 6 and 4 spokes per k-space frame were selected for
	single-slice, simultaneous three-slice and five-slice model-based brain
	reconstructions, respectively. For abdominal studies, 25 and 10 spokes per
	k-space frame were used for single-slice and simultaneous three-slice
	reconstructions.

	\subsection*{T1 Analysis}

	All quantitative T1 results are reported as mean $\pm$ standard deviation (SD).
	Regions-of-interest (ROIs) were carefully selected to minimize
	partial volume errors using the arrayShow \cite{sumpf2013arrayshow} tool in
	MATLAB (MathWorks, Natick, MA). For further analysis, relative difference maps
	$\big( \frac{|\text{T1}_\mathrm{estimate} -
		\text{T1}_\mathrm{ref}|}{|\text{T1}_\mathrm{ref}|}\times 100\% \big)$ and
	normalized relative errors ($\|\text{T1}_\mathrm{estimate} -
	\text{T1}_\mathrm{ref} \|_{2}/\|\text{T1}_\mathrm{ref}\|_{2}$) were calculated,
	respectively, with $\text{T1}_\mathrm{ref}$ the reference T1 map estimated from
	a single-slice acquisition \cite{Wang_Magn.Reson.Med._2018} and
	$\text{T1}_\mathrm{estimate}$ the T1 map reconstructed from the corresponding
	multi-slice acquisition. Moreover, the repeatability error between scans was calculated
	using $\sqrt{(\sum_{i=1}^{n_{s}}\text{T1}^{2}_{\text{diff}}(i))/n_{s}}$, with
	$\text{T1}_{\text{diff}}(i)$ the T1 difference between different measurements
	and $n_{s}$ the number of subjects. In addition, synthetic images are
	computed for all inversion times and this image series is then
	converted into a movie showing the recovery of the longitudinal magnetization
	after the inversion.

	\section*{Results}
	
	\label{sec:results}
	The proposed method was first validated in a phantom study using
	the SMS golden-angle acquisition and the developed SMS model-based
	reconstruction. Figure 2 (a) shows T1 maps of three slices reconstructed with
	the proposed method together with an IR spin-echo reference and the T1 maps
	estimated from single-slice IR radial acquisition
	\cite{Wang_Magn.Reson.Med._2018} of the center slice. Visual inspection reveals
	that the proposed method can completely disentangle the superposed slices and
	all multi-slice T1 maps are in good agreement with the references. These
	findings are confirmed by the T1 values of the ROIs in the center slice in
	Figure 2 (b), where preservation of good precision (low standard deviation) of
	the SMS T1 mapping method is also observed. Quantitative results for T1 maps
	from the other two slices are shown in the Supporting Information Table S1, which confirms
	good T1 accuracy and precision of these two slices as well.

	Figure 3 demonstrates the effect of the minimum regularization parameter
	$\alpha_{\text{min}}$ used in the SMS model-based reconstruction with a
	multi-slice factor of five. Low values of $\alpha_{\text{min}}$ result in
	increased noise in the T1 maps, while high values lead to blurring. A similar
	effect can be seen in the synthesized image series presented in the Supporting Information Video S1. 
	A value of $\alpha_{\text{min}}$ = 0.001 was chosen to balance 
	noise reduction and preservation of image details. This value was used for all
	multi-slice brain reconstructions. Similarly, $\alpha_{\text{min}}$ = 0.00025
	was used for all multi-slice abdominal reconstructions.

	Figure 4 (a) compares the center-slice T1 maps of a human brain for different
	multi-slice acquisitions and using model-based reconstruction for a
	multi-slice factor of three. Both SMS methods produce T1 maps with less noise
	than the conventional spoke-interleaved multi-slice method. Further, the
	SMS T1 values are closer to the single-slice reference, as seen
	in the relative difference maps as well as the normalized errors. In line
	with the three-slice results, Figure 4 (b) shows for a multi-slice factor
	of five that both SMS T1 maps have better SNR and are closer to the reference
	than the conventional multi-slice method. For both acceleration factors, the
	SMS golden-angle method further helps to reduce artifacts in the border areas
	and has less quantitative errors than the SMS aligned method. The T1 values
	for white- and gray-matter ROIs in Figure 5 confirm the above findings: Apart
	from similar mean T1 values among all multi-slice methods, both SMS approaches
	produce T1 values with higher precision than the spoke-interleaved
	multi-slice method. In addition, all SMS brain T1 values are in close agreement
	with the single-slice reference as well as literature values
	\cite{wansapura1999nmr,preibisch2009influence}. Note the 5-slice SMS aligned T1 map 
	looks slightly less noisier than the 3-slice SMS aligned one but has a higher normalized error. 
	This may be due to the fact that streaking artifacts start to appear in the border areas 
	of the 5-slice SMS aligned T1 map.

	Figure 6 (a) presents the center-slice abdominal T1 maps for different
	three-slice acquisitions using model-based reconstructions. Similar to the brain
	results, both SMS T1 maps have better SNR than the spoke-interleaved multi-slice result. 
	The SMS golden-angle method further helps to reduce streaking artifacts on the T1 maps. 
	The ROI-analyzed quantitative liver T1 values in Figure 6(b) confirm good accuracy of 
	all multi-slice methods in comparison to the single-slice T1 map with a best quantitative 
	precision achieved by the SMS golden-angle method.
	
	Figure 7 (a) depicts all T1 maps estimated from a five-slice SMS golden-angle
	acquisition as well as the corresponding single-slice T1 maps. All five-slice
	T1 maps are visually in good agreement with the corresponding single-slice maps.
	This is also observed for the other five subjects as shown in the
	Supporting Information Figure S1, indicating that the combination of the SMS
	golden-angle acquisition with SMS model-based reconstructions can
	be used to simultaneously acquire T1 maps for five slices of a
	human brain in good quality within four seconds.
	A similar comparison is presented for simultaneous
	three-slice abdominal T1 mapping in Figure 7(b) and Supporting Information Figure S2. In
	this case, the simultaneous three-slice datasets were acquired within a single
	four-second breathhold. Again, good agreement is reached between the
	simultaneous three-slice abdominal T1 maps and the single-slice maps. These results suggest an acceleration
	factor of three can be used for abdominal T1 mapping with the proposed method.
	Images for all inversion times were synthesized for the
	five-slice brain and three-slice liver studies and
	are presented in the Supporting Information Videos S2 and S3, respectively.

	Figure 8 and Figure 9 show T1 values of all six subjects for the
	two repetitive scans in both SMS brain and liver studies. The
	small T1 difference between the repetitive scans demonstrates good
	intra-subject repeatability of the proposed method. The repeatability 
	errors are: Frontal white matter: 13 ms (1.8$\%$ of the mean), occipital white
	matter: 4 ms (0.5$\%$ of the mean), frontal gray matter: 15 ms (1.1$\%$ of the
	mean), occipital  gray matter: 18 ms (1.3$\%$ of the mean) and liver: 4 ms
	(0.5$\%$ of the mean). Figure 9 additionally confirms good quantitative 
	agreement between single-slice and SMS liver T1 values for all subjects.

	Figure 10 illustrates the use of the proposed SMS T1 mapping technique
	for the acquisition of a whole-brain T1 map consisting of 25
	contiguous slices with a resolution of 0.75 $\times$ 0.75 $\times$ 5
	$\text{mm}^{3}$. These datasets were acquired using five consecutive 
	five-slice SMS acquisitions with each having slice distance of 25 mm.
	As a sufficient delay is necessary between the non-selective inversions, the total acquisition
	took around one minute: Four seconds for each five-slice SMS acquisition with
	a waiting period of 10 seconds in between to ensure full recovery of the
	magnetization.

	\section*{Discussion}
	\label{sec:Discussion}

	This work describes a fast multi-slice T1 mapping technique which combines
	simultaneous multi-slice excitations and single-shot IR radial FLASH with an
	extended nonlinear model-based reconstruction. The present method avoids any
	coil-calibration steps for SMS reconstruction and uses sparsity
	constraints to improve precision. Validation studies on a phantom and in
	six healthy subjects show that the combination of SMS golden-angle acquisition
	and the developed SMS model-based reconstruction technique
	can obtain high-resolution simultaneous five-slice brain T1 maps and
	simultaneous three-slice abdominal T1 maps with good accuracy, precision and
	repeatability within a single inversion recovery of four seconds.

	In comparison, to reconstruct a fully-sampled image at each time point while
	still keeping bin size small, the conventional IR Look-Locker technique needs
	to employ multiple inversions and  acquire complementary k-space data in each
	inversion to fulfill the Nyquist criterion.  With a bin size of 20 radial
	spokes (single-slice) and a 256 $\times$ 256 matrix, at least 
	$20 \approx (256 \times \pi /2) / 20$ inversions are needed. Thus, the 
	undersampling factor is 20 for model-based reconstruction in the
	single-shot single-slice method. With the extension to SMS described in the present
	work, the acceleration factor is further enhanced by an additional
	factor of 3-5. 
	The present method achieves 0.8 to 1.33 seconds per T1 map within a single
	inversion recovery and 2.8 to 4.67 seconds per map when taking into
	account a 10 second delay in a multi-slice protocol with several inversions.
	This compares well to other recent radial T1 mapping approaches 
	with IR Look-Locker based acquisitions
	recently described in the literature \cite{Maier2019,li2019rapid}.

	In contrast to many other quantitative SMS MRI techniques
	\cite{weingartner2017simultaneous,ye2017simultaneous,hilbert2019fast}, the
	proposed method integrates coil-sensitivity estimation into the model-based
	reconstruction framework, avoiding additional calibration scans, which 
	further reduces miscalibration errors in case of motion, especially
	when imaging moving organs such as abdomen or heart.
	Moreover, the proposed method dispenses with intermediate image
	reconstruction, enabling the reconstruction of high-quality T1 maps directly
	from k-space which does not require binning. This property is especially
	beneficial for T1 mapping of multiple slices at higher acceleration factors
	where other methods need to trade off image quality against accuracy
	when choosing the bin size \cite{wang2015single,deichmann2005fast}.
	The spoke-interleaved results in this
	study are noisier than the ones presented in \cite{Wang2018}. The main 
	reason might be that measurements in the previous work \cite{Wang2018}
	were done on a different scanner using a 64-channel head coil while this
	study employed a 20-channel head coil. Different parameter choices
	for sequence and reconstruction may also be a contributing factor.
	In this study, we used identical parameters as used in the proposed SMS
	model-based method to make a direct comparison possible.

	In the present work, we use a radial FLASH readout while
	other studies employ EPI or 3D spiral to achieve high-resolution
	brain mapping \cite{cohen2018optimized,cao2019fast}.
	While radial FLASH is less efficient in covering k-space,
	with short echo times it is much less sensitive to off-resonance
	effects than spiral or EPI trajectories
	and is not affected by geometric distortions. Due to the intrinsic
	oversampling of the k-space center, radial FLASH is also more robust
	to  motion, which will be beneficial in abdominal
	and cardiac applications. 
	
	An alternative to multi-slice T1 mapping is 3D imaging using a radial
	stack-of-stars sequence or other 3D sequences \cite{cao2019fast}. 3D
	sequences should provide even better SNR than 2D SMS imaging and allow
	for isotropic T1 mapping \cite{Maier2019, cao2019fast}. The proposed
	model-based reconstruction method is also applicable to 3D T1 mapping.
	However, 3D sequences need to employ multi-shot acquisitions, in which
	case data over the duration of a complete scan (usually in the order of
	minutes) have to be combined for parameter estimation. This makes 3D
	imaging much more sensitive to motion than their 2D SMS
	counterparts \cite{ye2017simultaneous}.

	For radial sampling, the use of a golden angle in the partition dimension has
	been shown to have better performance than the use of aligned spokes for
	undersampled stack-of-stars 3D volume MRI \cite{zhou2017golden} and 2D radial
	SMS parallel imaging using NLINV \cite{Rosenzweig_Magn.Reson.Med._2018}. In this
	work, we have adopted the same strategy for SMS model-based parameter
	mapping. Our results confirm a slightly better performance of this
	golden-angle strategy over the aligned case. Combination of such a sampling
	strategy and model-based reconstructions may be further exploited for
	single-shot SMS myocardial T1 mapping, where part of the IR data (e.g.,
	systolic data) would be discarded prior to model-based T1 estimations
	\cite{wang2019model}. The proposed method is also applicable to
	whole-liver imaging where we need to perform the single-shot
	acquisition multiple times, i.e., with multiple breathholds to achieve
	the desired volume coverage. As the delay time can be as short as 3
	seconds for liver studies \cite{chen2016rapid}, covering the entire liver
	with 30 slices using ten 4-second breathhold SMS-3 acquisitions interrupted
	by nine 3-second time gaps when the subject is allowed to
	breath freely required a minimum acquisition time of 67 seconds. With
	a longer delays of 10 seconds between breathholds for more patient
	convenience this scan could still be performed in two minutes.

	The non-selective inversion pulse used in this study is not optimal for
	whole-brain T1 mapping as it necessitates a delay time between
	successive SMS acquisitions. The 10-second delay time we employed is
	very high as usually 3-4 seconds should be enough for brain
	applications \cite{Maier2019}. Furthermore, it has been reported
	\cite{deichmann2005fast,li2019rapid} that there is little difference
	in T1 accuracy between non-selective and slab-selective inversion pulses
	for brain T1 mapping. With adapted excitation pulses and acquisition
	strategies, we expect to eliminate most of the delay
	time when combining a slab-selective inversion with the proposed
	simultaneous multi-slice T1 mapping technique for full brain
	applications in future studies.

	The other limitation of the proposed method is the long computation
	time, especially for model-based reconstruction of the SMS golden-angle
	datasets: All the data from multiple slices have to be held in memory
	simultaneously during iterations, which still prevents the use of GPUs.
	In principle, the alternative methods such as subspace methods have
	to deal with the same amount of data, but recent implementations
	use smart computational strategies to reduce the amount of memory
	and can then achieve very fast reconstruction \cite{mani2015fast, 
	Tamir_Magn.Reson.Med._2017,Uecker__2015}. The key
	difference between subspace and the nonlinear methods is that the
	latter does not need any approximations for the signal models, i.e., a
	minimal number of physical parameters could describe the desired MR
	signal precisely. In contrast, subspace methods have to approximate the MR
	signal using a few principal coefficients, which then leads to
	compromised accuracy or precision. While computation times reported in
	this work for the proposed model-based reconstruction on CPUs are still
	very high, newer generations of GPUs with larger memory will enable
	much faster reconstructions. Preliminary results obtained by using
	a smaller matrix size are very encouraging:
	When reducing the in-plane resolution from 0.75 $\times$ 0.75 mm$^{2}$
	(matrix size: 512 $\times$ 512) to 1.0 $\times$ 1.0 mm$^{2}$ (matrix
	size: 384 $\times$ 384), the SMS 5-slice model-based reconstructions
	could already run on a GPU, which then took only around 14 min per dataset.
	Similarly, the 1.0 $\times$ 1.0 mm$^{2}$ SMS 3-slice reconstruction
	took around 8 min per dataset.

	\section*{Conclusion}
	\label{sec:Conclusion}
	
	The proposed combination of simultaneous multi-slice excitations, single-shot
	IR radial FLASH, and calibrationless model-based reconstruction allows for
	efficient high-resolution multi-slice T1 mapping with good accuracy, precision
	and repeatability.


\section*{Conflict of Interest}

The authors declare no competing interests.

\section*{Data Availability Statement}
In the spirit of reproducible research, code to
reproduce the experiments is available on
\url{https://github.com/mrirecon/sms-T1-mapping}.
The raw k-space data used in this study can be downloaded from
\href{https://doi.org/10.5281/zenodo.3969809}{DOI:10.5281/zenodo.3969809}.

	\bibliographystyle{mrm}
	\bibliography{radiology}

\begin{thebibliography}{10}

\bibitem{margaret2012practical}
Cheng~HL, Stikov~N, Ghugre~NR, Wright~GA.
\newblock Practical medical applications of quantitative {MR} relaxometry.
\newblock J Magn Reson Imaging. 2012; 36:805--824.

\bibitem{kellman2014t1}
Kellman~P, Hansen~MS.
\newblock T1-mapping in the heart: accuracy and precision.
\newblock J Cardiovasc Magn Reson. 2014; 16:2.

\bibitem{Look_Rev.Sci.Instrum._1970}
Look~DC, Locker~DR.
\newblock {T}ime saving in measurement of {NMR} and {EPR} relaxation times.
\newblock Rev Sci Instrum. 1970; 41:250--251.

\bibitem{deichmann1992quantification}
Deichmann~R, Haase~A.
\newblock Quantification of {T}1 values by {SNAPSHOT}-{FLASH} {NMR} imaging.
\newblock J Magn Reson. 1992; 96:608--612.

\bibitem{Messroghli_Magn.Reson.Med._2004}
Messroghli~DR, Radjenovic~A, Kozerke~S, Higgins~DM, Sivananthan~MU, Ridgway~JP.
\newblock {M}odified {L}ook-{L}ocker {I}nversion recovery ({MOLLI}) for
  high-resolution {T}1 mapping of the heart.
\newblock Magn Reson Med. 2004; 52:141--146.

\bibitem{Block_IEEETrans.Med.Imaging_2009}
Block~KT, Uecker~M, Frahm~J.
\newblock {M}odel-{B}ased {I}terative {R}econstruction for {R}adial {F}ast
  {S}pin-{E}cho {MRI}.
\newblock IEEE Trans Med Imaging. 2009; 28:1759–1769.

\bibitem{Doneva_Magn.Reson.Med._2010}
Doneva~M, B{\"o}rnert~P, Eggers~H, Stehning~C, S{\'e}n{\'e}gas~J, Mertins~A.
\newblock Compressed sensing reconstruction for magnetic resonance parameter
  mapping.
\newblock Magn Reson Med. 2010; 64:1114--1120.

\bibitem{Petzschner_Magn.Reson.Med_2011}
Petzschner~FH, Ponce~IP, Blaimer~M, Jakob~PM, Breuer~FA.
\newblock Fast {MR} parameter mapping using k-t principal component analysis.
\newblock Magn Reson Med. 2011; 66:706--716.

\bibitem{huang2012t2}
Huang~C, Graff~CG, Clarkson~EW, Bilgin~A, Altbach~MI.
\newblock T2 mapping from highly undersampled data by reconstruction of
  principal component coefficient maps using compressed sensing.
\newblock Magn Reson Med. 2012; 67:1355--1366.

\bibitem{velikina2013accelerating}
Velikina~JV, Alexander~AL, Samsonov~A.
\newblock Accelerating {MR} parameter mapping using sparsity-promoting
  regularization in parametric dimension.
\newblock Magn Reson Med. 2013; 70:1263--1273.

\bibitem{zhang2015accelerating}
Zhang~T, Pauly~JM, Levesque~IR.
\newblock Accelerating parameter mapping with a locally low rank constraint.
\newblock Magn Reson Med. 2015; 73:655--661.

\bibitem{Tamir_Magn.Reson.Med._2017}
Tamir~JI, Uecker~M, Chen~W, Lai~P, Alley~MT, Vasanawala~SS, Lustig~M.
\newblock {T}2 shuffling: {S}harp, multicontrast, volumetric fast spin-echo
  imaging.
\newblock Magn Reson Med. 2017; 77:180--195.

\bibitem{feng2017compressed}
Feng~L, Benkert~T, Block~KT, Sodickson~DK, Otazo~R, Chandarana~H.
\newblock Compressed sensing for body {MRI}.
\newblock J Magn Reson Imaging. 2017; 45:966--987.

\bibitem{gensler2014myocardial}
Gensler~D, M{\"o}rchel~P, Fidler~F, Ritter~O, Quick~HH, Ladd~ME, Bauer~WR,
  Ertl~G, Jakob~PM, Nordbeck~P.
\newblock Myocardial {T}1 quantification by using an {ECG}-triggered radial
  single-shot inversion-recovery {MR} imaging sequence.
\newblock Radiology. 2014; 274:879--887.

\bibitem{wang2016high}
Wang~X, Joseph~AA, Kalentev~O, Merboldt~KD, Voit~D, Roeloffs~VB, van Zalk~M,
  Frahm~J.
\newblock High-resolution myocardial {T}1 mapping using single-shot inversion
  recovery fast low-angle shot {MRI} with radial undersampling and iterative
  reconstruction.
\newblock Br J Radiol. 2016; 89:20160255.

\bibitem{marty2018fast}
Marty~B, Coppa~B, Carlier~P.
\newblock Fast, precise, and accurate myocardial {T}1 mapping using a radial
  {MOLLI} sequence with {FLASH} readout.
\newblock Magn Reson Med. 2018; 79:1387--1398.

\bibitem{Fessler_IEEESignalProcess.Mag._2010}
Fessler~JA.
\newblock Model-based image reconstruction for {MRI}.
\newblock IEEE Signal Process Mag. 2010; 27:81--89.

\bibitem{Sumpf_J.Magn.Reson.Imaging_2011}
Sumpf~TJ, Uecker~M, Boretius~S, Frahm~J.
\newblock {M}odel-based nonlinear inverse reconstruction for {T}2~{m}apping
  using highly undersampled spin-echo {MRI}.
\newblock J Magn Reson Imaging. 2011; 34:420–428.

\bibitem{tran2013model}
Tran{-}Gia~J, St{\"a}b~D, Wech~T, Hahn~D, K{\"o}stler~H.
\newblock Model-based acceleration of parameter mapping ({MAP}) for saturation
  prepared radially acquired data.
\newblock Magn Reson Med. 2013; 70:1524--1534.

\bibitem{zhao2014model}
Zhao~B, Lam~F, Liang~ZP.
\newblock Model-based {MR} parameter mapping with sparsity constraints:
  parameter estimation and performance bounds.
\newblock IEEE Trans Med Imaging. 2014; 33:1832--1844.

\bibitem{peng2014exploiting}
Peng~X, Liu~X, Zheng~H, Liang~D.
\newblock Exploiting parameter sparsity in model-based reconstruction to
  accelerate proton density and {T}2 mapping.
\newblock Med Eng Phys. 2014; 36:1428--1435.

\bibitem{knoll2015model}
Knoll~F, Raya~JG, Halloran~RO, Baete~S, Sigmund~E, Bammer~R, Block~T, Otazo~R,
  Sodickson~DK.
\newblock A model-based reconstruction for undersampled radial spin-echo {DTI}
  with variational penalties on the diffusion tensor.
\newblock NMR Biomed. 2015; 28:353--366.

\bibitem{Volkert_Int.J.Imag.Syst.Tech._2016}
Roeloffs~V, Wang~X, Sumpf~TJ, Untenberger~M, Voit~D, Frahm~J.
\newblock Model-based reconstruction for {T}1 mapping using single-shot
  inversion-recovery radial {FLASH}.
\newblock Int J Imag Syst Tech. 2016; 26:254--263.

\bibitem{zhao2016maximum}
Zhao~B, Setsompop~K, Ye~H, Cauley~SF, Wald~LL.
\newblock Maximum likelihood reconstruction for magnetic resonance
  fingerprinting.
\newblock IEEE Trans Med Imaging. 2016; 35:1812--1823.

\bibitem{Wang_Magn.Reson.Med._2018}
Wang~X, Roeloffs~V, Klosowski~J, Tan~Z, Voit~D, Uecker~M, Frahm~J.
\newblock Model-based {T}1 mapping with sparsity constraints using single-shot
  inversion-recovery radial {FLASH}.
\newblock Magn Reson Med. 2018; 79:730--740.

\bibitem{Maier2019}
Maier~O, Schoormans~J, Schloegl~M, Strijkers~GJ, Lesch~A, Benkert~T, Block~T,
  Coolen~BF, Bredies~K, Stollberger~R.
\newblock Rapid {T}1 quantification from high resolution 3{D} data with
  model-based reconstruction.
\newblock Magn Reson Med. 2019; 81:2072--2089.

\bibitem{moon2013myocardial}
Moon~JC, Messroghli~DR, Kellman~P, Piechnik~SK, Robson~MD, Ugander~M,
  Gatehouse~PD, Arai~AE, Friedrich~MG, Neubauer~S et~al.
\newblock Myocardial {T}1 mapping and extracellular volume quantification: a
  {S}ociety for {C}ardiovascular {M}agnetic {R}esonance ({SCMR}) and {CMR}
  {W}orking {G}roup of the {E}uropean {S}ociety of {C}ardiology consensus
  statement.
\newblock J Cardiovasc Magn Reson. 2013; 15:92.

\bibitem{weingartner2017simultaneous}
Weing{\"a}rtner~S, Moeller~S, Schmitter~S, Auerbach~E, Kellman~P, Shenoy~C,
  Ak{\c{c}}akaya~M.
\newblock Simultaneous multislice imaging for native myocardial {T}1 mapping:
  {I}mproved spatial coverage in a single breath-hold.
\newblock Magn Reson Med. 2017; 78:462--471.

\bibitem{shah2001new}
Shah~NJ, Zaitsev~M, Steinhoff~S, Zilles~K.
\newblock A new method for fast multislice {T}1 mapping.
\newblock Neuroimage. 2001; 14:1175--1185.

\bibitem{deichmann2005fast}
Deichmann~R.
\newblock Fast high-resolution {T}1 mapping of the human brain.
\newblock Magn Reson Med. 2005; 54:20--27.

\bibitem{Wang2018}
Wang~X, Voit~D, Roeloffs~V, Uecker~M, Frahm~J.
\newblock Fast interleaved multislice {T}1 mapping: Model-based reconstruction
  of single-shot inversion-recovery radial {FLASH}.
\newblock Comput Math Methods Med. 2018; 2018.

\bibitem{li2019rapid}
Li~Z, Bilgin~A, Johnson~K, Galons~JP, Vedantham~S, Martin~DR, Altbach~MI.
\newblock Rapid high-resolution {T}1 mapping using a highly accelerated radial
  steady-state free-precession technique.
\newblock J Magn Reson Imaging. 2019; 49:239--252.

\bibitem{Barth_Magn.Reson.Med._2016}
Barth~M, Breuer~F, Koopmans~PJ, Norris~DG, Poser~BA.
\newblock {S}imultaneous multislice ({SMS}) imaging techniques.
\newblock Magn Reson Med. 2016; 75:63--81.

\bibitem{ye2017simultaneous}
Ye~H, Cauley~SF, Gagoski~B, Bilgic~B, Ma~D, Jiang~Y, Du~YP, Griswold~MA,
  Wald~LL, Setsompop~K.
\newblock Simultaneous multislice magnetic resonance fingerprinting
  ({SMS}-{MRF}) with direct-spiral slice-{GRAPPA}(ds-{SG}) reconstruction.
\newblock Magn Reson Med. 2017; 77:1966--1974.

\bibitem{hilbert2019fast}
Hilbert~T, Schulz~J, Marques~JP, Thiran~JP, Krueger~G, Norris~DG, Kober~T.
\newblock Fast model-based {T}2 mapping using {SAR}-reduced simultaneous
  multislice excitation.
\newblock Magn Reson Med. 2019; 82:2090--2103.

\bibitem{wundrak2016golden}
Wundrak~S, Paul~J, Ulrici~J, Hell~E, Geibel~MA, Bernhardt~P, Rottbauer~W,
  Rasche~V.
\newblock Golden ratio sparse {MRI} using tiny golden angles.
\newblock Magn Reson Med. 2016; 75:2372--2378.

\bibitem{zhou2017golden}
Zhou~Z, Han~F, Yan~L, Wang~DJ, Hu~P.
\newblock Golden-ratio rotated stack-of-stars acquisition for improved
  volumetric {MRI}.
\newblock Magn Reson Med. 2017; 78:2290--2298.

\bibitem{Rosenzweig_Magn.Reson.Med._2018}
Rosenzweig~S, Holme~HCM, Wilke~RN, Voit~D, Frahm~J, Uecker~M.
\newblock Simultaneous multi‐slice {MRI} using cartesian and radial {FLASH}
  and regularized nonlinear inversion: {SMS}-{NLINV}.
\newblock Magn Reson Med. 2018; 79:2057--2066.

\bibitem{Uecker_Magn.Reson.Med._2008}
Uecker~M, Hohage~T, Block~KT, Frahm~J.
\newblock {I}mage reconstruction by regularized nonlinear inversion - joint
  estimation of coil sensitivities and image content.
\newblock Magn Reson Med. 2008; 60:674--682.

\bibitem{Beck_IEEETrans.ImageProcessing_2009}
Beck~A, Teboulle~M.
\newblock {F}ast {G}radient-{B}ased {A}lgorithms for {C}onstrained {T}otal
  {V}ariation {I}mage {D}enoising and {D}eblurring {P}roblems.
\newblock IEEE Trans Image Process. 2009; 18:2419--2434.

\bibitem{Barral_Magn.Reson.Med._2010}
Barral~JK, Gudmundson~E, Stikov~N, Etezadi‐Amoli~M, Stoica~P, Nishimura~DG.
\newblock {A} robust methodology for in vivo {T}1 mapping.
\newblock Magn Reson Med. 2010; 64:1057--1067.

\bibitem{Uecker__2015}
Uecker~M, Ong~F, Tamir~JI, Bahri~D, Virtue~P, Cheng~JY, Zhang~T, Lustig~M.
\newblock {B}erkeley advanced reconstruction toolbox.
\newblock { In:} Proc. Intl. Soc. Mag. Reson. Med. 23, Toronto, 2015.  p. 2486.

\bibitem{Block_J.KoreanSoc.Magn.Reson.Med._2014}
Block~KT, Chandarana~H, Milla~S, Bruno~M, Mulholland~T, Fatterpekar~G,
  Hagiwara~M, Grimm~R, Geppert~C, Kiefer~B.
\newblock {T}owards routine clinical use of radial stack-of-stars 3{D}
  gradient-echo sequences for reducing motion sensitivity.
\newblock J. Korean Soc. Magn. Reson. Med. 2014; 18:87--106.

\bibitem{Block__2011}
Block~KT, Uecker~M.
\newblock Simple method for adaptive gradient-delay compensation in radial
  {MRI}.
\newblock { In:} Proc. Intl. Soc. Mag. Reson. Med. 19, Montreal, 2011.  p.
  2816.

\bibitem{Wajer__2001}
Wajer~FTAW, Pruessmann~KP.
\newblock {M}ajor speedup of reconstruction for sensitivity encoding with
  arbitrary trajectories.
\newblock { In:} Proc. Intl. Soc. Mag. Reson. Med. 9, Glasgow, 2001.  p. 0767.

\bibitem{Uecker_Magn.Reson.Med._2010}
Uecker~M, Zhang~S, Frahm~J.
\newblock {N}onlinear inverse reconstruction for real-time {MRI} of the human
  heart using undersampled radial {FLASH}.
\newblock Magn Reson Med. 2010; 63:1456–1462.

\bibitem{sumpf2013arrayshow}
Sumpf~T, Unterberger~M.
\newblock arrayshow: a guide to an open source matlab tool for complex mri data
  analysis.
\newblock { In:} Proc. Intl. Soc. Mag. Reson. Med. 21, Salt Lake City, 2013.
  p. 2719.

\bibitem{wansapura1999nmr}
Wansapura~JP, Holland~SK, Dunn~RS, BallJr~WS.
\newblock {NMR} relaxation times in the human brain at 3.0 tesla.
\newblock J Magn Reson Imaging. 1999; 9:531--538.

\bibitem{preibisch2009influence}
Preibisch~C, Deichmann~R.
\newblock Influence of {RF} spoiling on the stability and accuracy of {T}1
  mapping based on spoiled {FLASH} with varying flip angles.
\newblock Magn Reson Med. 2009; 61:125--135.

\bibitem{wang2015single}
Wang~X, Roeloffs~V, Merboldt~KD, Voit~D, Sch{\"a}tz~S, Frahm~J.
\newblock Single-shot multi-slice {T}1 mapping at high spatial
  resolution--inversion-recovery {FLASH} with radial undersampling and
  iterative reconstruction.
\newblock Open Med Imaging J. 2015; 9.

\bibitem{cohen2018optimized}
Cohen~O, Polimeni~JR.
\newblock Optimized inversion-time schedules for quantitative {T}1 measurements
  based on high-resolution multi-inversion {EPI}.
\newblock Magn Reson Med. 2018; 79:2101--2112.

\bibitem{cao2019fast}
Cao~X, Ye~H, Liao~C, Li~Q, He~H, Zhong~J.
\newblock Fast 3{D} brain {MR} fingerprinting based on multi-axis spiral
  projection trajectory.
\newblock Magn Reson Med. 2019; 82:289--301.

\bibitem{wang2019model}
Wang~X, Kohler~F, UnterbergBuchwald~C, Lotz~J, Frahm~J, Uecker~M.
\newblock Model-based myocardial {T}1 mapping with sparsity constraints using
  single-shot inversion-recovery radial {FLASH} cardiovascular magnetic
  resonance.
\newblock J Cardiovasc Magn Reson. 2019; 21:60.

\bibitem{chen2016rapid}
Chen~Y, Lee~GR, Aandal~G, Badve~C, Wright~KL, Griswold~MA, Seiberlich~N,
  Gulani~V.
\newblock Rapid volumetric {T}1 mapping of the abdomen using three-dimensional
  through-time spiral {GRAPPA}.
\newblock Magnetic resonance in medicine 2016; 75:1457--1465.

\bibitem{mani2015fast}
Mani~M, Jacob~M, Magnotta~V, Zhong~J.
\newblock Fast iterative algorithm for the reconstruction of multishot
  non-cartesian diffusion data.
\newblock Magn Reson Med. 2015; 74:1086--1094.

\end{thebibliography}
	
	\clearpage

\begin{figure}
	\centering
	\includegraphics[width=1.0\textwidth]{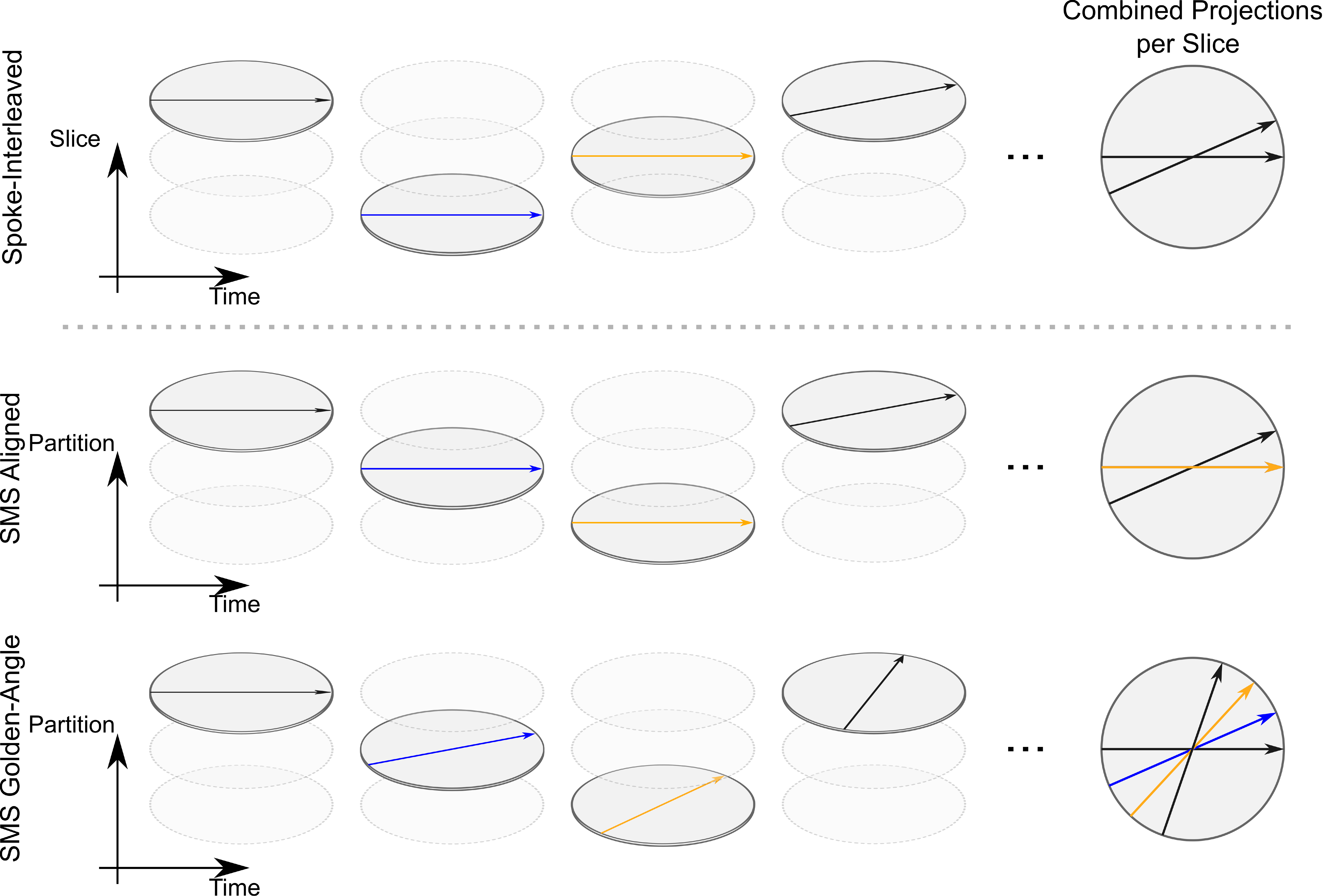}
	\caption{Single-shot IR radial multi-slice data acquisition schemes.
		(Top) Conventional spoke-interleaved multi-slice acquisition scheme. (Middle
		and bottom) Radial SMS with spokes distributed following aligned and
		golden-angle distribution in the partition dimension, respectively. Note
		that the longitudinal axis represents slice direction for the conventional
		spoke-interleaved multi-slice scheme and the partition dimension for the SMS
		acquisitions. For a certain slice, the SMS golden-angle acquisition scheme could
		cover more k-space within a given readout time after inversion.}
\end{figure}

\begin{figure}
	\centering
	\includegraphics[width=1.0\textwidth]{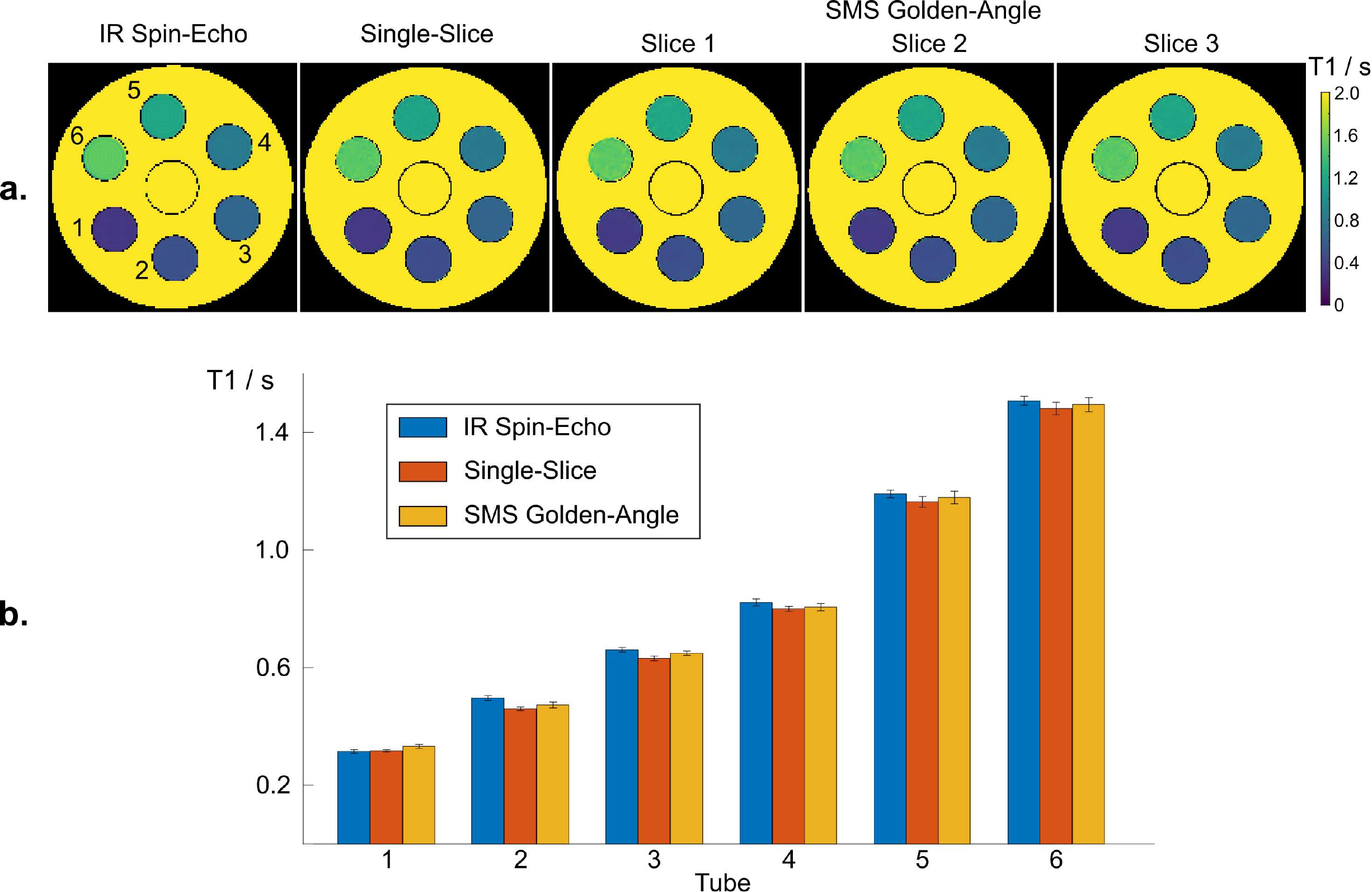}
	\caption{\textbf{a}. T1 maps for three simultaneously acquired slices 
		for a phantom using a three-slice SMS golden-angle IR FLASH acquisition
		and model-based reconstruction in comparison to a single-slice IR spin-echo 
		reference method (center slice) and a single-slice IR FLASH acquisition
		using model-based reconstruction.
		\textbf{b}. Quantitative T1 values (mean and standard deviation) within ROIs 
		of the six phantom tubes for the center-slice T1 maps of all three methods.
	}
\end{figure}

\begin{figure}
	\centering
	\includegraphics[width=1.0\textwidth]{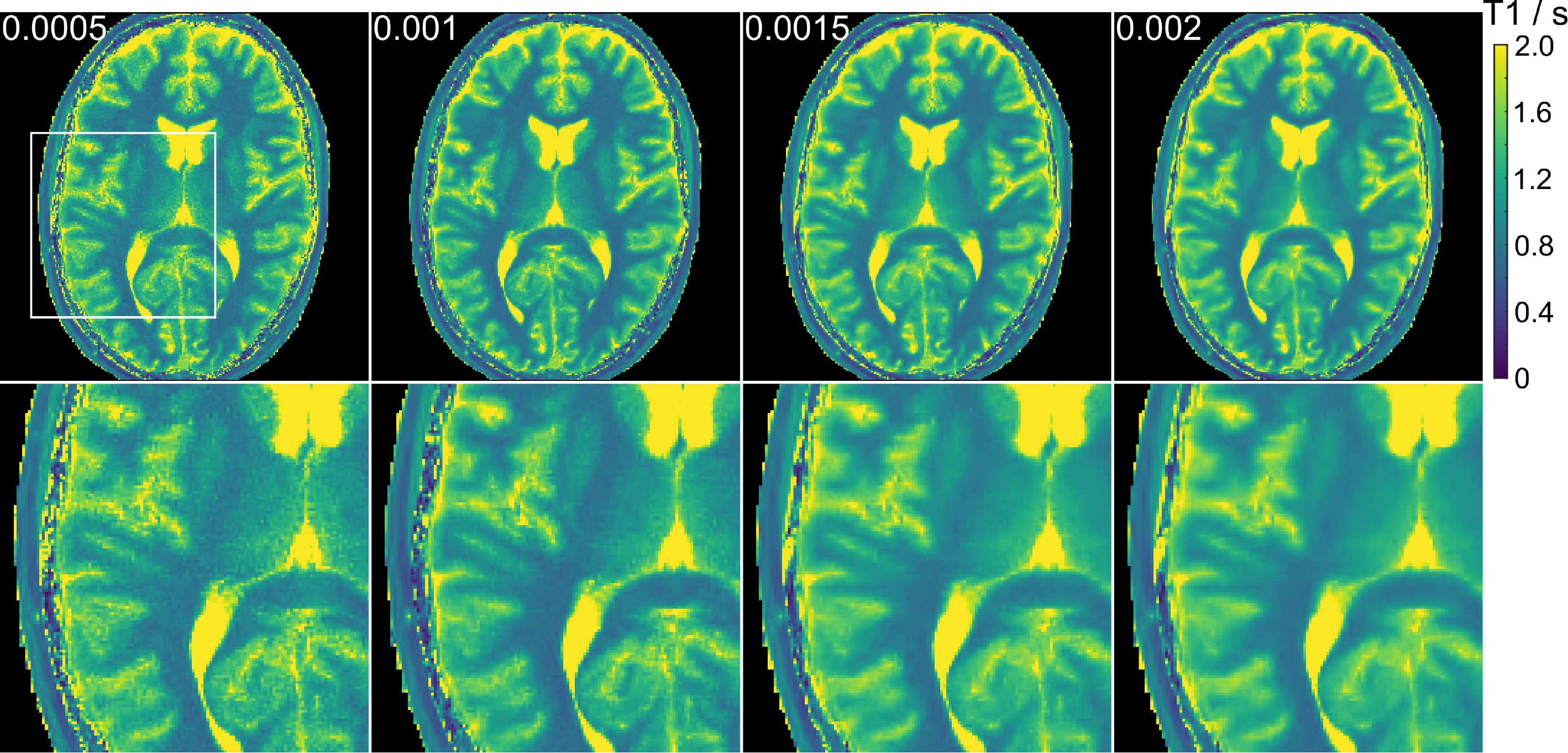}
	\caption{\textbf{a}. Five-slice SMS brain T1 maps (center
		slice) obtained using model-based reconstruction for different
		choices of the minimum regularization parameters $\alpha_{\text{min}}$.
		A value $\alpha_{\text{min}}$ = 0.001 is used for all
		multi-slice brain studies.
	}
\end{figure}

\begin{figure}
	\centering
	\includegraphics[width=0.85\textwidth]{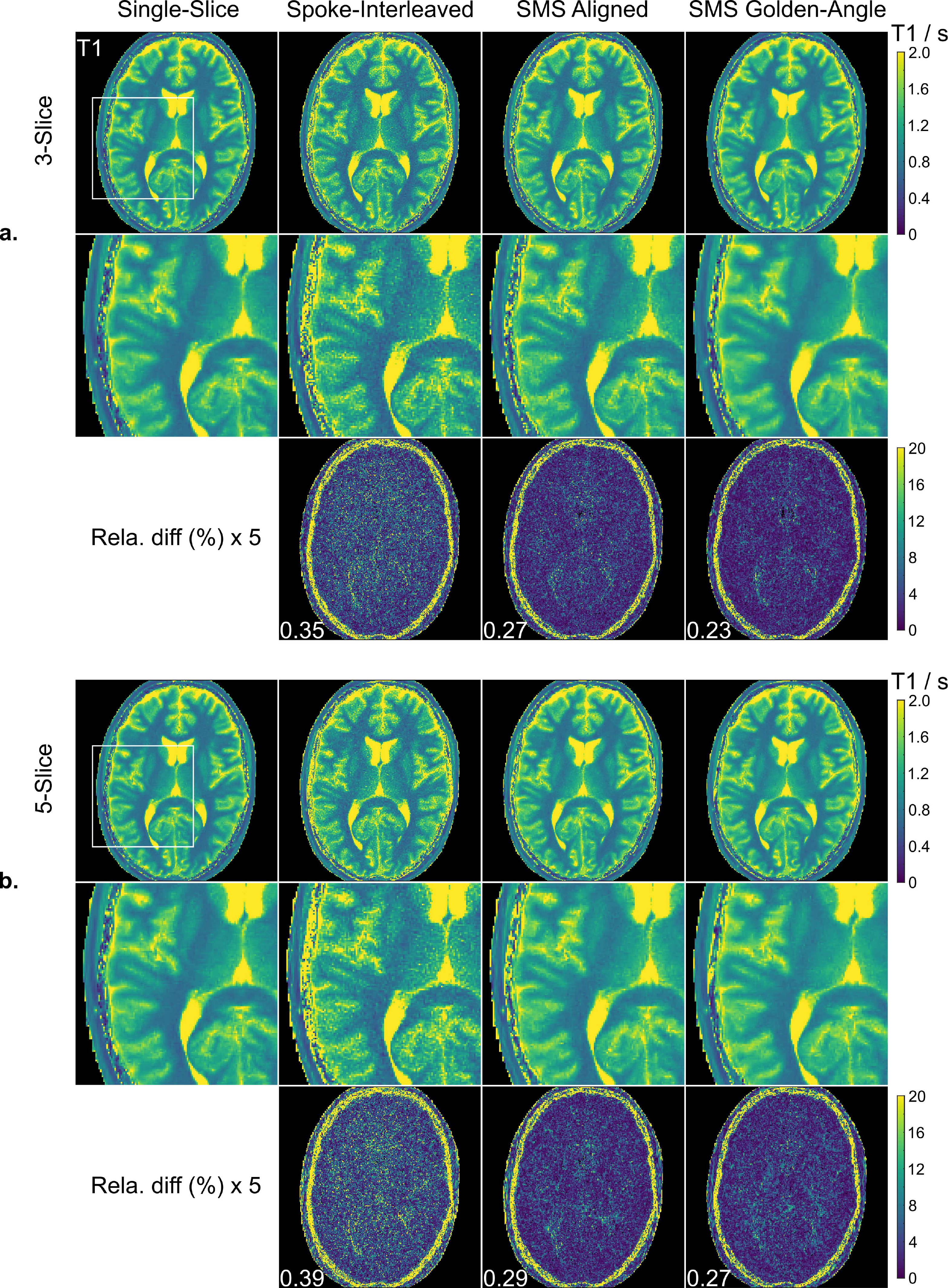}
	\caption{\textbf{a}. (Top) Center-slice of three-slice brain T1 maps
		from different multi-slice acquisitions using model-based reconstructions with
		(middle) magnified T1 regions and (bottom) their relative difference maps
		($\times$ 5) to the single-slice reference T1 map. \textbf{b}. Similar
		comparisons at a multi-slice factor of five. Normalized errors are
		presented at the bottom left or all relative-difference T1 maps.
	}
\end{figure}	

\begin{figure}
	\centering
	\includegraphics[width=1.0\textwidth]{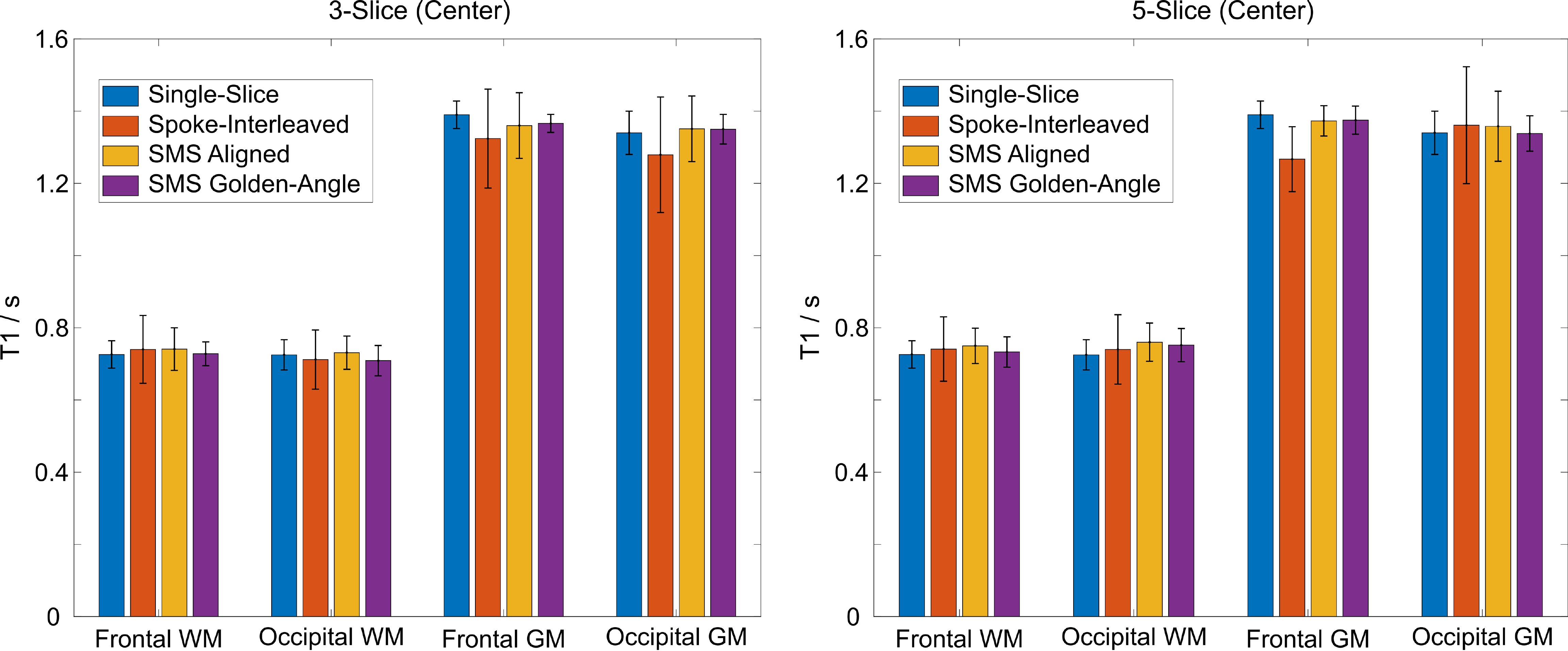}
	\caption{Quantitative T1 values (mean and standard deviation) within
		ROIs that were manually drawn into the frontal white matter (WM), occipital WM,
		frontal gray matter (GM) and occipital GM of all T1 maps in Figure 4.
	}
\end{figure}

\begin{figure}
	\centering
	\includegraphics[width=1.0\textwidth]{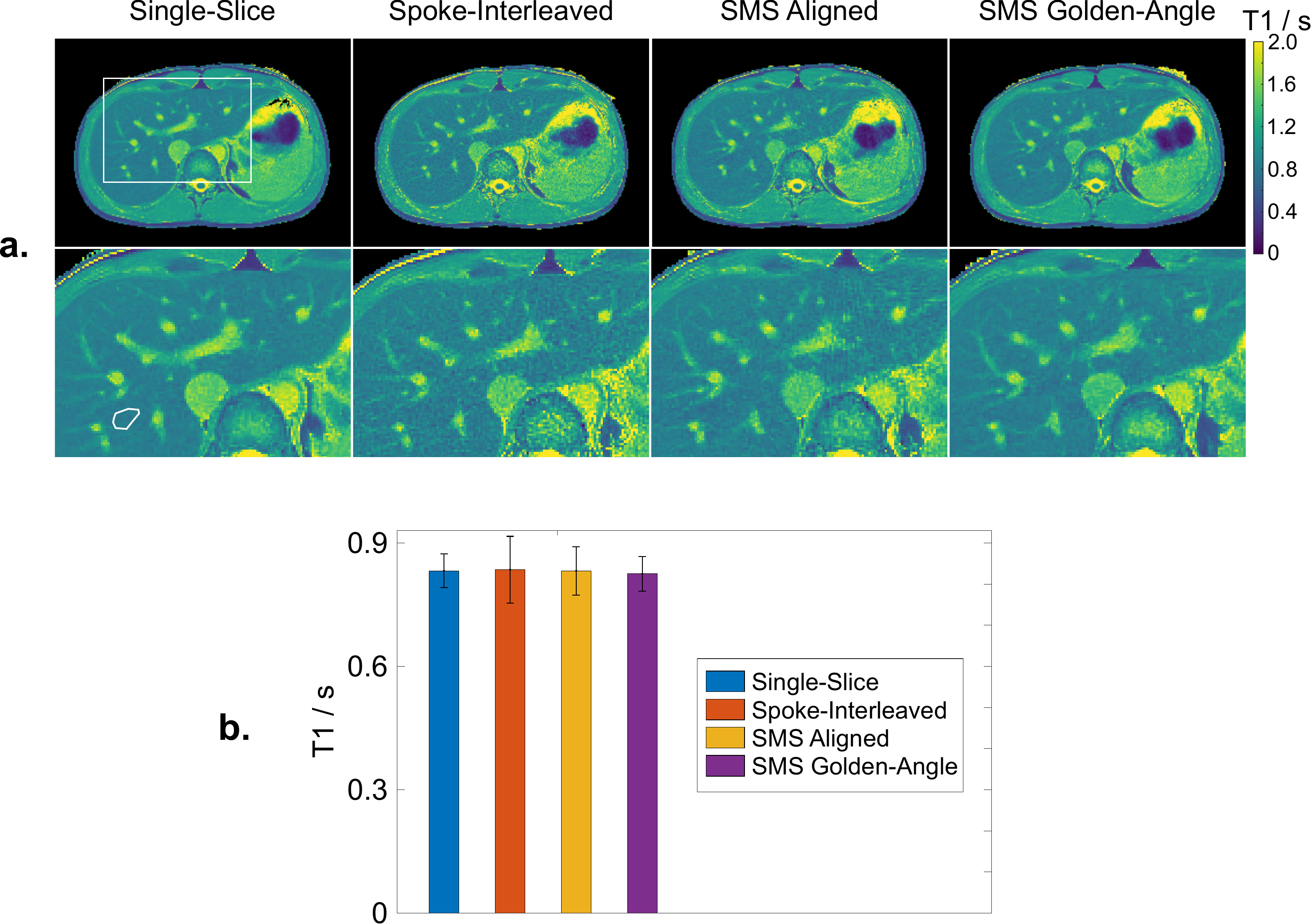}
	\caption{ \textbf{a}. (Top) Center-slice of three-slice abdominal T1 maps
		from different multi-slice acquisitions using model-based reconstructions with
		(bottom) magnified T1 regions. \textbf{b}. Quantitative T1 values (mean and standard deviation) within
		ROIs that were manually drawn into the liver region as indicated in \textbf{a}.	}
\end{figure}

\begin{figure}
	\centering
	\includegraphics[width=1.0\textwidth]{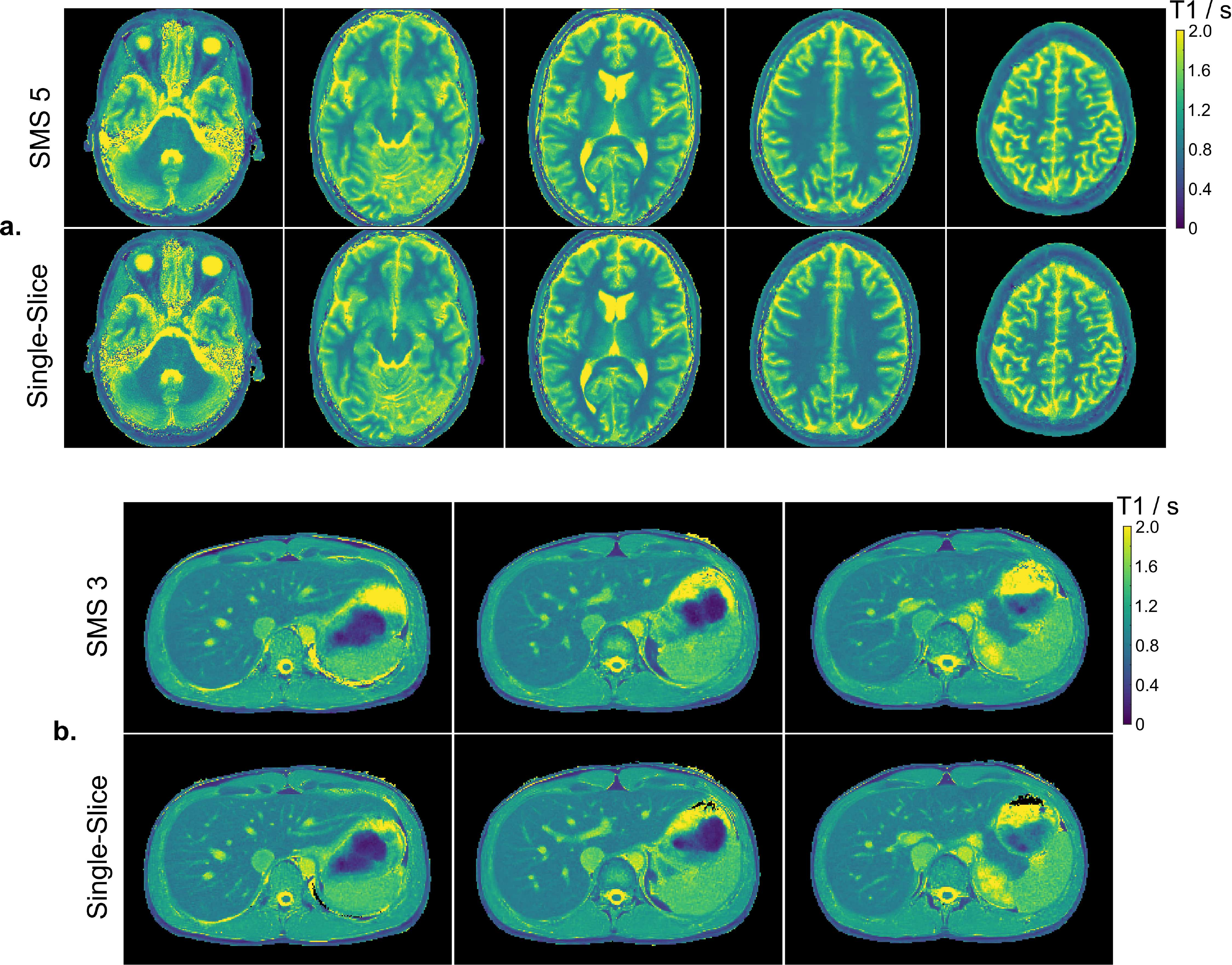}
	\caption{ \textbf{a}. (Top) T1 maps of five simultaneously acquired 
		slices of a human brain obtained using the SMS golden-angle acquisition and the
		proposed model-based reconstruction technique. (Bottom) The corresponding
		single-slice T1 maps. \textbf{b}. Similar comparisons for simultaneous
		three-slice abdominal T1 mapping. The abdominal acquisition was
		performed within a single breathhold of four seconds.
	}
\end{figure}

\begin{figure}
	\centering
	\includegraphics[width=1.0\textwidth]{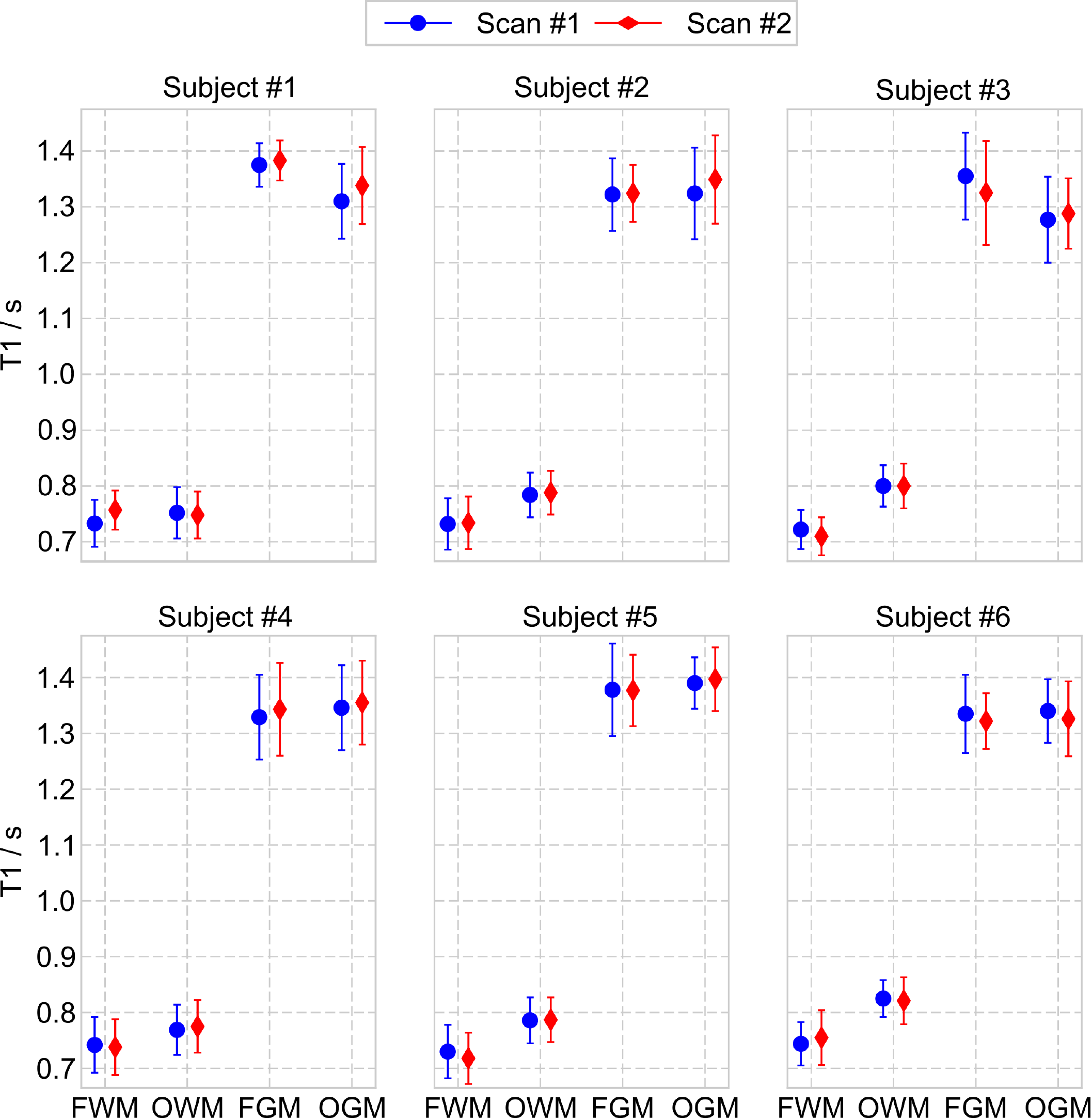}
	\caption{ \textbf{a}.  Quantitative brain T1 values (s) in the four ROIs of the center slice when using the five-slice SMS T1 methods for
		all six subjects and two repetitive scans. FWM - frontal white matter, OWM -
		occipital white matter, FGM - frontal gray matter and OGM - occipital gray
		matter.
	}
\end{figure}

\begin{figure}
	\centering
	\includegraphics[width=1.0\textwidth]{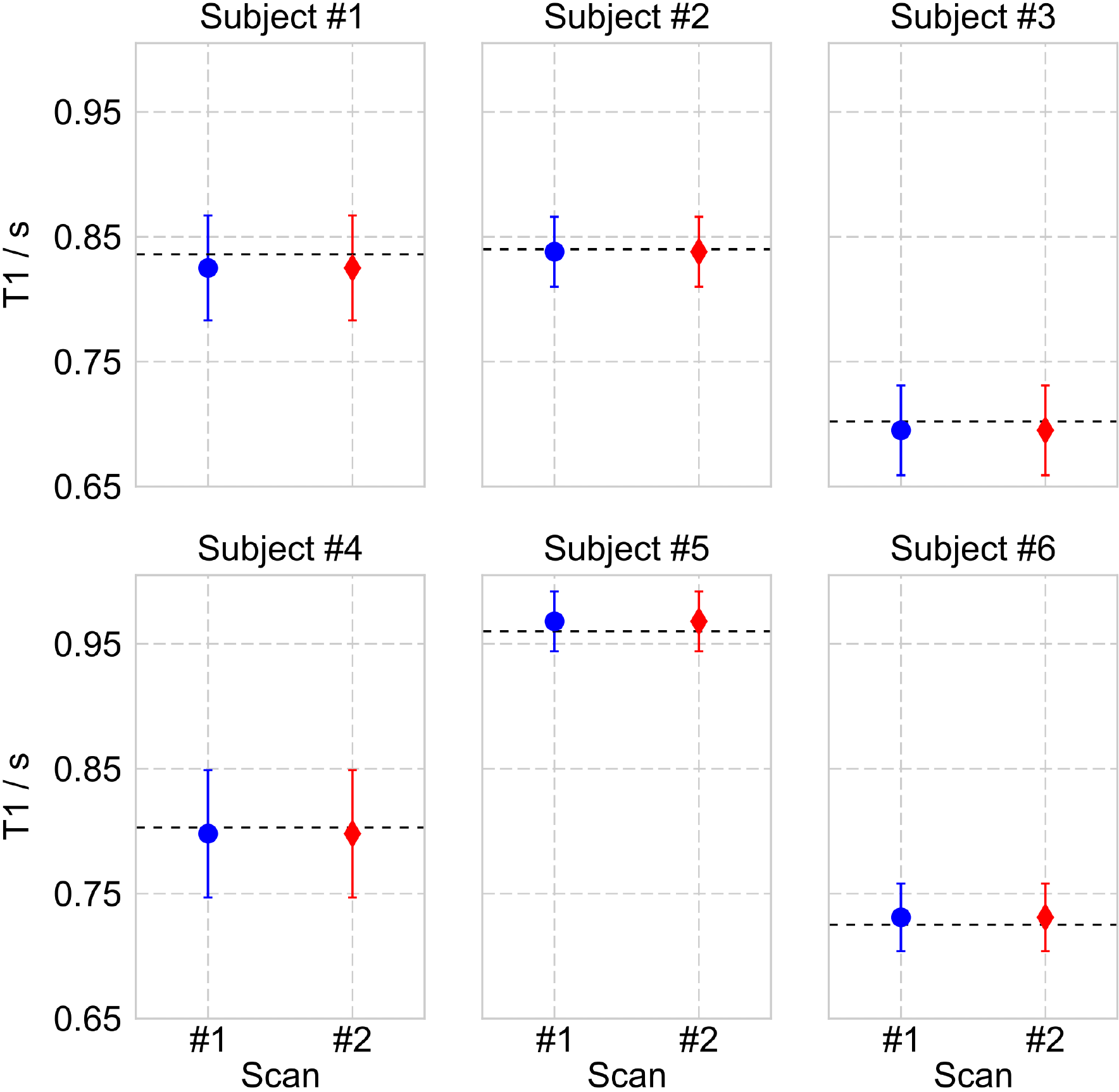}
	\caption{ Quantitative liver T1 values (s) of the center slice
		when using the three-slice SMS T1 methods for all six subjects and two repetitive scans.
		Note that dashed black lines represent the corresponding liver T1 values from
		single-slice acquisitions.
	}
\end{figure}

\begin{figure}
	\centering
	\includegraphics[width=1.0\textwidth]{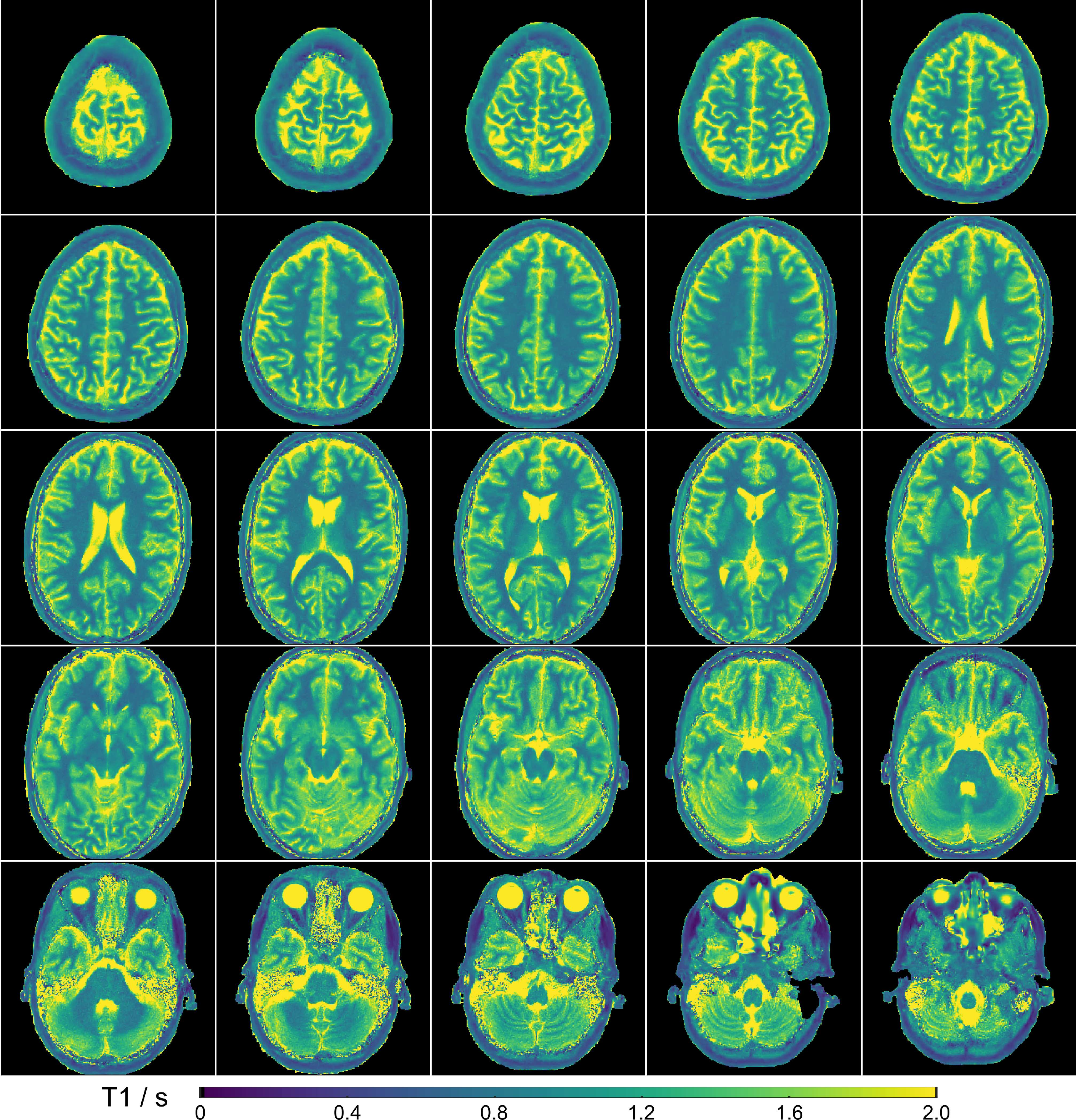}
	\caption{ Full brain T1 map consisting of 25
		contiguous slices at a resolution of 0.75 $\times$ 0.75 $\times$ 5 mm$^{3}$.
		These 25 T1 maps were acquired using five consecutive five-slice
		SMS acquisitions with a slice distance of 25 mm each.
	}
\end{figure}

	\section*{Supplementary Materials}
	
	\textbf{Supporting Information Table S1.} T1 relaxation times (ms, mean $\pm$ SD) for the experimental phantom in Figure 2.
	
	\textbf{Supporting Information Figure S1.} 
	T1 maps for five simultaneously acquired slices of
	the human brain and the corresponding single-slice 
	T1 maps for five subjects.
	
	\textbf{Supporting Information Figure S2.} 
	T1 maps for three simultaneously acquired abdominal
	slices and the corresponding single-slice T1 maps for
	five subjects. 
	
	\textbf{Supporting Information Video S1.}
	Synthesized image series representing recovery
	of the longitudinal magnetization after inversion
	for the five-slice SMS brain study
	(center-slice) using model-based reconstruction with
	different minimum regularization parameters,
	i.e., $\alpha_{\text{min}}$ = 0.0005, 0.001, 0.0015, 0.002, from left to right, respectively.
	
	\textbf{Supporting Information Video S2.}
	Synthesized image series representing recovery
	of the longitudinal magnetization after inversion
	of five simultaneously slices of a human
	brain when using a bin size of 82 ms. The
	acquisition time for all five slices is four
	seconds.
	
	\textbf{Supporting Information Video S3.}
	Synthesized image series representing recovery
	of the longitudinal magnetization after inversion
	for three simultaneously acquired slices of
	a human liver using a bin size of 81 ms.
	The acquisition time for all three slices is
	four seconds.

\end{document}